\documentclass[12pt]{article}

\usepackage{amssymb}
\usepackage{amsmath}
\usepackage{latexsym}

\catcode`@=11
\renewcommand{\section}{\@startsection{section}{1}{0pt}{\medskipamount}
{\medskipamount}{\large\bf}} \numberwithin{equation}{section}
\catcode`@=12

\def\beq{\begin{eqnarray}}    
\def\eeq{\end{eqnarray}}      

\def\ln{\,\mbox{ln}\,}                  

\def\pa{\partial}                       

\def\={\ =\ }

\begin{document}
\begin{center}

{\Large\bf A systematic study of finite BRST-BV transformations
within $W$-$X$ formulation of the standard and the $Sp(2)$-extended
field-antifield formalism}

\vspace{18mm}

{\Large Igor A. Batalin$^{(a,c)}\footnote{E-mail: batalin@lpi.ru}$\;,
Klaus Bering$^{(b)}\footnote{E-mail: bering@physics.muni.cz}$\;,
Peter M. Lavrov$^{(c, d)}\footnote{E-mail: lavrov@tspu.edu.ru}$}

\vspace{8mm}

\noindent ${{}^{(a)}}$
{\em P.N. Lebedev Physical Institute,\\
Leninsky Prospect \ 53, 119 991 Moscow, Russia}

\noindent ${{}^{(b)}}$
{\em Masaryk University, Faculty of Science,\\
Kotlarska 2, 611 37 Brno, Czech Republic}

\noindent  ${{}^{(c)}} ${\em
Tomsk State Pedagogical University,\\
Kievskaya St.\ 60, 634061 Tomsk, Russia}

\noindent  ${{}^{(d)}} ${\em
National Research Tomsk State University,\\
Lenin Av.\ 36, 634050 Tomsk, Russia}

\vspace{20mm}

\begin{abstract}
\noindent Finite BRST-BV transformations are studied systematically
within the $W$-$X$ formulation of the standard and the $Sp(2)$-extended
field-antifield formalism. The finite BRST-BV transformations are
introduced by formulating a new version of the Lie equations.
The corresponding finite change of the gauge-fixing master action $X$ and
the corresponding Ward identity are derived.
\end{abstract}

\end{center}

\vfill


\vspace{5mm}
\section{Introduction}

In recent papers \cite{LL,FDBRST-1,FDBRST-2,FDBRST-3,FDBRST-4,FDBRST-5},
finite BRST transformations have been
studied systematically both in the Hamiltonian and Lagrangian
formalism in their standard and $Sp(2)$-extended versions
\cite{FV,BVhf,BV,BV2,BLTh1,BLTh2,BLTh3,Sp,BLTl1,Hu,BLTl2,BLTl3}. The
so-called $W$-$X$ formulation \cite{BT1,BT2,BT3,BM,BMS,BBD1,BBD2,BB,BB1}
is known as the most symmetric form of
the Lagrangian field-antifield formalism. Dynamical gauge-generating
master action $W$ serves as a deformation to the original
action of the theory. On the other hand, gauge-fixing master
action $X$ serves just as to eliminate the antifield variables. It is
remarkable that these complementary master actions $W$ and $X$ do satisfy
a set of quantum master equations transposed to each other.

In the present paper we study systematically finite BRST-BV
transformations within the $W$-$X$ formulation both in the standard and
$Sp(2)$-extended field-antifield formalism. We introduce these
transformations by formulating the respective Lie equations. Among
other things, we derive in this way the effective change in the
gauge-fixing master action $X$, as induced by the finite BRST-BV
transformation defined.

\vspace{5mm}
\section{$W$-$X$ formulation to the standard field-antifield
formalism}

Let $z^{ A }$ be the complete set of the variables necessary within
the standard field-antifield formalism
\beq
\label{I1}
z^{ A } = \{ \Phi^{ \alpha };
\Phi^{*}_{ \alpha } \}, 
\eeq
whose Grassmann parities are
\beq
\label{I2}
\varepsilon( z^{ A } )
= \{ \varepsilon_{ \alpha }; \varepsilon_{ \alpha } + 1\}. 
\eeq
We denote the respective $z^{ A }$-derivatives as
\beq
\label{I3}
\pa_{ A } = \{ \pa_{ \alpha }; \pa^{ \alpha }_{*} \}. 
\eeq
Let $Z$ be the partition function
\beq
\label{I4}
Z  =  \int  Dz D\lambda  \exp \left\{\frac{i}{ \hbar } \big[ W + X \big]
\right\} , 
\eeq
where $\lambda^{\alpha}$ are Lagrange multipliers for gauge-fixing with
Grassmann parity
\beq
\label{I5}
\varepsilon( \lambda^{ \alpha } )  = \varepsilon_{ \alpha } + 1. 
\eeq
In the partition function (\ref{I4}), the dynamical gauge-generating
master action $W$ and the gauge-fixing master action $X$ are defined to
satisfy the respective quantum master equations,
\beq
\label{I6}
\left(\Delta  \exp\left\{ \frac{i}{\hbar } W \right\}  \right) =  0
\quad\Leftrightarrow\quad
\frac{1}{2} ( W , W ) = i\hbar (\Delta W),        
\eeq
\beq
\label{I7}
\left(\Delta  \exp\left\{\frac{i}{\hbar } X \right\}\right)  =  0
\quad\Leftrightarrow\quad
\frac{1}{2} ( X, X )  =i  \hbar (\Delta X).          
\eeq
In the above quantum master equations (\ref{I6}) and (\ref{I7}),
the $\Delta$ and $( \;, \;)$ are the standard nilpotent odd Laplacian
\beq
\label{I8}
\Delta  =  \pa_{ \alpha } \pa^{ \alpha }_{*} (-1)^{ \varepsilon_{ \alpha } }, 
\eeq
and the standard antibracket
\beq
\label{I9}
( f, g )  = (-1)^{ \varepsilon_{ f }}  [ [ \Delta, f  ], g  ] 1
= f \overleftarrow{\pa }_{ \alpha }~\overrightarrow{\pa }^{\alpha}_{*}  g  -
( f \leftrightarrow g ) (-1)^{ ( \varepsilon_{ f } + 1 ) ( \varepsilon_{ g} + 1 ) }, 
\eeq
respectively. These formulae (\ref{I8}) and (\ref{I9}) tell us that the
anticanonical pairs $(\Phi^{ \alpha }; \Phi^{*}_{ \alpha } )$ serve as Darboux
coordinates on the flat field-antifield phase space with measure density
$\rho=1$ and no odd scalar curvature $\nu_{\rho}=0$.

At $\hbar = 0$, $\Phi^{*}_{ \alpha } = 0$, the $W$-action coincides with the
original action of the theory. As to the $X$-action, it can be chosen in
the form related to the gauge-fixing Fermion $\Psi( \Phi )$,
\beq
\label{I10}
X =  ( \Phi^{*}_{ \alpha }  -  \Psi( \Phi )\overleftarrow{\pa }_{ \alpha } )
\lambda^{ \alpha }
=  \Phi^{*}_{ \alpha }  \lambda^{ \alpha } -\Psi( \Phi ) \overleftarrow{d }, 
\eeq
where
\beq
\label{I11}
\overleftarrow{d } = \overleftarrow{\pa }_{\alpha}~ \lambda^{ \alpha } 
\eeq
is a nilpotent Fermionic differential that acts from the right.

In the integrand of the path integral (\ref{I4}),
consider now the following infinitesimal BRST-BV transformation
\beq
\label{I12}
\delta z^{ A }
= -\mu (Y, z^{ A } )  -  \frac{\hbar}{i}(\mu, z^{ A } )
= -\frac{\hbar}{i} y^{-1}(y\mu, z^{ A }) ,
\eeq
where we have defined for later convenience
\beq
\label{I12b}
 Y~:=~X-W,\qquad
y~:=~  \exp\left\{ \frac{i}{\hbar}Y \right\},
\eeq
and where $\mu(z)$ is an infinitesimal Fermionic function with
$\varepsilon(\mu)=1$.
The Jacobian of the infinitesimal BRST-BV transformation (\ref{I12}) has the
form
\beq
\label{I13}
\ln J = (-1)^{\varepsilon_{A}} (\partial_{A} \delta z^{ A })
=  (Y,  \mu )  +  2  (\Delta Y)  \mu  +
2 \frac{\hbar}{i}  (\Delta \mu). 
\eeq
The complete action in the partition function (\ref{I4}) transforms as
\beq
\label{I14}
\delta [ W + X ]  =  \left[ - ( W,W )  +  ( X, X ) \right] \mu
+  \frac{\hbar}{i} (  W + X, \mu  ).         
\eeq
Due to the quantum master equations (\ref{I6}) and (\ref{I7}), we then have
from Eqs.\ (\ref{I13}) and (\ref{I14}) that
\beq
\label{I15}
\frac{i}{\hbar}\delta [ W + X ] +  \ln J =  2 (\sigma( X )\mu), 
\eeq
where $\sigma( X )$ is a quantum BRST generator
\beq
\label{I16}
(\sigma( X )f)   =   (X, f)  +  \frac{\hbar}{i} (\Delta f)   . 
\eeq
The Eq.\ (\ref{I15}) tells us that the BRST transformation (\ref{I12}) induces
the following variation
\beq
\label{I17}
\delta X  =  2\frac{\hbar}{i} (\sigma( X )\mu). 
\eeq
to the $X$-action in the integrand of the path integral (\ref{I4}).
We conclude that the partition function (\ref{I4}) and the quantum master
equation (\ref{I7}) for $X$ are both stable under the infinitesimal variation
(\ref{I17}).

Next let $t$ be a Bosonic parameter. It is natural to define a one-parameter
subgroup $t\mapsto \overline{z}^{ A }(t)$ of finite BRST-BV transformations
by the Lie equation\footnote{For an arbitrary function $f=f(z)$, we use the
shorthand notation
$\overline{f}=f(\overline{z})=\left(e^{ t \mathcal{ H } }f\right)$
for the corresponding function with shifted arguments.}
\beq
\label{I18}
\frac{d\overline{z}^{A}}{dt} = \overline{\mathcal{H}}^{A},
\qquad \overline{z}^{A} |_{  t=0 }   =  z^{ A }. 
\eeq
where
\beq
\label{I30}
\mathcal{ H }  =  \mathcal{ H }^{A} \partial_{A} =
- \mu \; {\rm ad}( Y )  -   \frac{\hbar}{i} \; {\rm ad}( \mu )
=  -\frac{\hbar}{i} y^{-1} {\rm ad} ( y \mu),
\qquad
{\rm ad}( f ) g :=  ( f , g ), \quad
\eeq
is the corresponding vector field with components
\beq
\label{I30b}
\mathcal{ H }^{A} := -\mu (Y, z^{ A } )  -  \frac{\hbar}{i}(\mu, z^{ A } )
= -\frac{\hbar}{i} y^{-1}(y\mu, z^{ A }) .
\eeq
Note that $\mu(z)$ is now an arbitrary finite Fermionic function.
In other words, the Lie equation (\ref{I18}) is
\beq
\label{I30c}
\frac{d\overline{z}^{A}}{dt}
=(\overline{\mathcal{H}}\overline{z}^{A}), \qquad
\overline{\mathcal{ H }}
:=  \overline{\mathcal{ H }}^{A} \underline{\partial}_{A}
\eeq
with solution
\beq
\label{I30d}
\overline{z}^{A}=\left( e^{ t \mathcal{ H } } z^{ A } \right).
\eeq
Recall that the antibracket for any Fermion $F=y\mu$ with itself is zero:
$( F,F ) = 0$. This fact yields a conservation law
\beq
\label{I18a}
\frac{ d (\overline{y} \overline{\mu}) }{ dt }
= \frac{d\overline{z}^{A}}{dt}~
\underline{\partial}_{A}(\overline{y} \overline{\mu})
= -\frac{\hbar}{i} \overline{y}^{-1}\overline{(y\mu,z^{A})}~
\underline{\partial}_{A}(\overline{y} \overline{\mu})
=  -\frac{\hbar}{i} \overline{y}^{-1}\overline{(y\mu,y\mu)} =  0,  
\eeq
so that the following invariance property holds
\beq
\label{I18b}
 \overline{y}\overline{\mu } = y  \mu. 
\eeq
The Jacobian of these transformations satisfies the following equation
\beq
\label{I19}
\frac{d\ln J}{dt}  = \overline{\mathrm{div}\mathcal{H}},
\qquad \mathrm{div}  \mathcal{H}
:=  (-1)^{\varepsilon_{A}} \partial_{A} \mathcal{H}^{A}
= (Y, \mu  ) + 2  (\Delta Y ) \mu
+ 2 \frac{\hbar}{i}(\Delta \mu ) . 
\eeq
The transformed complete action satisfies the equation
\beq
\nonumber
\label{I20}
\frac{d}{dt} \left[\overline{W } + \overline{X} \right]
&=& \frac{d\overline{z}^{A}}{dt}~ \underline{\partial}_{A}
\left[\overline{W } + \overline{X} \right]
= \left[-\overline{\mu} \overline{(Y, z^{ A } )}
 -  \frac{\hbar}{i}\overline{(\mu, z^{ A } )} \right]
\underline{\partial}_{A} \left[\overline{W } + \overline{X} \right]  \\
&=& \left[ - \overline{( W , W )}
+  \overline{( X , X )}   \right] \overline{\mu }   +
\frac{\hbar}{i}  \overline{(W +X , \mu  )}. 
\eeq
Due to the transformed master equations (\ref{I6}) and (\ref{I7}), it follows
that
\beq
\label{I21}
\frac{d}{dt} \left[ \overline{W }  + \overline{X }
+\frac{\hbar}{i} \ln J\; \right] = \frac{\hbar}{i}\overline{a},  
\eeq
where we have defined for later convenience
\beq
\label{Ia}
 a:=  2 (\sigma  (  X  ) \mu) .
\eeq
By integrating Eq.\ (\ref{I21}) within $0 \leq t \leq 1$, we get
\beq
\label{I22}
 \overline{ W }  + \overline{X } +  \frac{\hbar}{i} \ln J
=   W  +  X  + \frac{\hbar}{i} A,   
\eeq
where we have defined the average
\beq
\label{I33}\label{I35}
 A :=   \int_{0}^{1} \! dt~ \overline{a }
=  \int_{0}^{1} \! dt~\left(e^{ t \mathcal{ H } }a\right)
= \left(E( \mathcal{ H } )a \right).
\eeq
Here $E$ is the function
\beq
\label{I36}
E( x )  = \int_{0}^{1} \! dt~e^{tx} = \frac{ \exp\{ x \} - 1 }{x}.   
\eeq
The Eq.\ (\ref{I22}) shows the finite effective change in $X$  induced by the
finite transformation $z^{ i }\rightarrow \overline{ z}^{ i }$
in the partition function (\ref{I4}).

Now consider the left-hand side $\overline{\mathcal{Y} }$ of the
transformed quantum master equation (\ref{I7}), where
\beq
\label{I23}
\mathcal{Y} :=  \frac{1}{2} (X,X ) +\frac{\hbar}{i}(\Delta X).   
\eeq
We have the following Cauchy initial value problem
\beq
\label{I24}
\frac{d\overline{\mathcal{Y}}}{dt}
= (\overline{\mathcal{H}}~\overline{\mathcal{Y}})
\quad\wedge\quad
\overline{\mathcal{Y} }|_{ t = 0 }=0
\quad\Rightarrow\quad\overline{\mathcal{Y} } \equiv 0
\eeq
for arbitrary $t$.

Thereby, we have confirmed that the quantum master equation (\ref{I7}) is
stable under the finite BRST-BV transformation generated by Eq.\
(\ref{I18}). Of course, the general expression (\ref{I4}) itself is stable
under the same transformation, as well.

At this point we would like to investigate the quantum master equation
\beq
\label{I25}
\left(\Delta \exp\left\{ \frac{i}{\hbar} X' \right\}\right)  =  0,     
\eeq
where we have denoted the new gauge-fixing master action,
\beq
\label{I26}
X'  =  X  + \frac{\hbar}{i}A.     
\eeq
Eq.\ (\ref{I25}) is equivalent to
\beq
\label{I34}
(\sigma( X ) \exp\{ A \})  =  0  \quad\Leftrightarrow\quad
\frac{\hbar}{2i} ( A, A ) + (\sigma( X ) A)  = 0.           
\eeq
The exponential $\exp\{ A \}$ rewrites in the form
\beq
\label{IA}
\exp\{ A \} =  e^{\left(E( \mathcal{ H } )a \right)}
= \left( \exp\{ \mathcal{ H } +a \}1\right)
= \left( \exp\{  H  + 2  [ \;\!\sigma( X ), \mu ] \}1\right),
\eeq
where we have defined the first-order operator
\beq
\label{I28}
H  := \frac{\hbar}{i}  y\; {\rm ad}( y^{-1} \mu)
= - \mu \;\! {\rm ad}(Y )  +  \frac{\hbar}{i} \;\! {\rm ad}( \mu )
=\mathcal{ H }+ 2 \frac{\hbar}{i}  {\rm ad}( \mu )  , 
\eeq
and used the formula
\beq
\label{I29}
[ \sigma( X ), f ] = ( \sigma( X ) f ) +(-1)^{\varepsilon_{f}} 
\frac{\hbar}{i}\;\! {\rm ad}( f )
\eeq
for a function $f$. Hence Eqs. (\ref{I25})/(\ref{I34}) is equivalent to
\beq
\label{I27}
\left[ \;\! \sigma( X ),~  \exp\left\{  H  +  2  [ \;\!\sigma( X ), \mu ]
 \;\!\right\}  \;\!\right]
  1  =  0.  
\eeq
In general, it looks as if Eq.\ (\ref{I25})/(\ref{I34})/(\ref{I27})
serves as a condition for finite field-dependent
parameter $\mu( z )$. This equation is certainly satisfied with arbitrary
infinitesimal $\mu( z ) \;\rightarrow \; 0$, to the first order in that.
We do not know if the same situation holds for arbitrary finite $\mu( z )$, as
Eq.\ (\ref{I25})/(\ref{I34})/(\ref{I27}) is rather complicated in the
general case.
Also, there is a potential obstacle that the dynamical master action $W$
actually enters that equation. Thus, being finite parameter $\mu( z )$
restricted in its field-dependence, that circumstance would be a crucial
specific feature of the $W$-$X$ formulation.

One can proceed from a solution $A$ to the quantum master equation (\ref{I34}).
If we ignore Eqs. (\ref{Ia}) and (\ref{I35}), then the quantum master equation
(\ref{I34}) knows nothing about the parameter $\mu$. Moreover, $A$ serves as an
external source in the left-hand side of the Eq.\ (\ref{I35}).
The right-hand side of Eq.\ (\ref{I35}) knows about parameter $\mu$ via
its explicit appearance in Eqs.\ (\ref{I30}) and (\ref{I35}).
Thereby, the aspects related to the quantum master equation (\ref{I34})
by itself, and to the parameter $\mu$, are separated naturally.
From this point of view, it sounds not so plausible that the
Eq.\ (\ref{I35}) could allow for finite arbitrary parameter $\mu( z )$.
If one rescales parameters in Eqs.\ (\ref{I34}) and (\ref{I35}),
\beq
\label{I37}
\mu \; \rightarrow\; \varepsilon \;\! \mu , 
\eeq
with $\varepsilon$ being a Boson parameter, and then expands $\mu$ and $A$
in formal power series,
\beq
\label{I38}
\mu   =  \mu_{0}  +  \varepsilon  \mu_{1}  +  \ldots ,   
\eeq
\beq
\label{I39}
A   =   A_{0}  +  \varepsilon A_{1} + \ldots ,  
\eeq
one gets to the first order in  $\varepsilon$
\beq
\label{I40}
0  =  A_{0},  
\eeq
\beq
\label{I41}
2 (\sigma( X )\;\! \mu_{0})  =  A_{1}, 
\eeq
\beq
\label{I42}
(\sigma( X )\;\! A_{1})  =  0,  
\eeq
so that $\mu_{0}$ remains arbitrary to that order. However, to the second
order in $\varepsilon$, one has
\beq
\label{I43}
2 (\sigma( X )\;\! \mu_{1}) + ( \mathcal{ H }(\mu_{0})
\;\!\sigma( X )\;\! \mu_{0})  = A_{2}, 
\eeq
\beq
\label{I44}
 \frac{\hbar}{2i}( A_{1}, A_{1} )  +  (\sigma( X )\;\! A_{2})  =  0, 
\eeq
so that $\mu_{1}$ remains arbitrary to that order, while (\ref{I44})
restricts $\mu_{0}$,
\beq
\label{I45}
(\sigma( X )\;\! H( \mu_{0} )\;\! \sigma( X )\;\! \mu_{0} ) =  0, 
\eeq
with $H( \mu_{0} )$  being the operator (\ref{I28}) as taken at $\mu = \mu_{0}$.
To the third order in $\varepsilon$,  the $\mu_{2}$  remains arbitrary,
while the $\mu_{1}$ is restricted to satisfy the condition
\beq
\nonumber
&&(\sigma( X )\;\! H( \mu_{0} )\;\! \sigma( X ) \mu_{1})  +
(\sigma( X )\;\! H( \mu_{1} )\;\! \sigma( X ) \mu_{0})   \\
&&+
\left(\frac{1}{3}\;\! \sigma( X ) \left(  H( \mu_{0} )  +
2  [\;\! \sigma( X ), \mu_{0} ]  \right)^{2}  \sigma ( X ) \mu_{0}\right)=0 .
\label{I46}
\eeq
The same situation holds to higher orders in $\varepsilon$:  to each
subsequent order, the respective coefficient in $\mu$ remains arbitrary, while
the preceding coefficient in  $\mu$ becomes restricted. Of course, it looks
rather difficult to estimate on being such a strange procedure "convergent"
to infinite order in $\varepsilon$.

It may look a bit strange that the operator $H$ from Eq.\ (\ref{I28}) appears
in Eqs.\ (\ref{I45}) and (\ref{I46}) while $\mathcal{ H }$
from Eq.\ (\ref{I30}) enters Eq.\ (\ref{I35}). In fact, one could,
in principle, proceed directly from the Eq.\ (\ref{I27}) formulated via the
operator $H$ from the very beginning. Then one could use the
Eq.\ (\ref{I27}), together with the properties
\beq
\label{I47}
(\sigma( X )1)  =  0, \quad    (H1)  =  0, 
\eeq
to derive the equations (\ref{I45}), (\ref{I46}).  Also, notice that there
is an implication
\beq
\label{I48}
( \sigma( X ) )^{2}  =  0  \;\Rightarrow \;
( \sigma( X )\; \mathcal{ O}\; \sigma( X ))^{2}  =  0,       
\eeq
with $\mathcal{ O }$ being any operator.

Finally we present a simple general argument, based on
the integration by parts, that the partition function (\ref{I4}) is independent
of finite arbitrariness in a solution to the gauge-fixing master action $X$,
\beq
\label{I49}
\exp\left\{ \frac{i}{\hbar} X' \right\}
&=&\left(
\exp\left\{[ \Delta, \mu ]\right\} \exp\left\{\frac{i}{\hbar} X \right\}
\right) \\
&=&  \exp\left\{ \frac{i}{\hbar} X \right\}  +
\left(\Delta\;\!\mu \;\! E( [ \Delta, \mu ] )
\exp \left\{ \frac{i}{\hbar} X \right\}\right),  
\label{I50}
\eeq
where  $\mu$ is any finite Fermionic operator and the function $E( x )$ is
defined in Eq.\ (\ref{I36}). By substituting
Eq.\ (\ref{I50}) into Eq.\ (\ref{I4}) with $X'$ standing for $X$, and then
integrating by parts with Eq.\ (\ref{I6}) taken into account, one observes that
the second term in the right-hand side in Eq.\ (\ref{I50}) does not contribute
to the integral (\ref{I4}). Thereby, the integral (\ref{I4}) with $X'$ standing
for $X$ reduces to the case of initial $X$ standing for itself. Thus, the
partition function is independent of a particular representative of the
class (\ref{I49}).

\vspace{5mm}
\section{Ward identities in the standard $W$-$X$ formulation}

Let $J_{ A }$ be external sources to the variables $z^{ A }$; then the
integral (\ref{I4}) generalizes to the generating functional,
\beq
\label{W1}
Z [ J ]  =  \int \! Dz D\lambda
\exp\left\{ \frac{i}{\hbar} [\;\! W + X + J_{ A } z^{ A }\;\! ]\right\}.
\eeq
Arbitrary variation $\delta z^{ A }$ yields the equations of motion,
\beq
\label{W2}
\left\langle\pa_{ B } ( W + X )\right\rangle_{J}  +
J_{ B } (-1)^{ \varepsilon_{ B } }   =  0,    
\eeq
where  $\langle \ldots \rangle_{J}$ is the source-dependent mean-value
\beq
\label{W3}
\langle \ldots  \rangle_{J}  =  \frac{1}{Z[J]}
\int Dz D\lambda  ( \ldots  )
\exp\left\{ \frac{i}{\hbar} [\;\! W + X + J_{ A } z^{ A }\;\! ] \right\}.
\eeq
It follows from Eq.\ (\ref{W2}) that
\beq
\label{W4}
\left \langle J_{ A }\; \omega^{ A B }\;
\pa_{ B } ( W  + X )\right\rangle_{J} +
J_{ A }\; \omega^{ A B }\; J_{ B } (-1)^{ \varepsilon_{ B } }   = 0,   
\eeq
where
\beq
\label{W5}
\omega^{ A B }  =  ( z^{ A },  z^{ B } ) =  {\rm const}     
\eeq
is the fundamental invertible antibracket. In Eq.\ (\ref{W1}), the BRST-BV
variation (\ref{I12}) yields
\beq
\label{W6}
\left\langle J_{ A } \;\omega^{ A B }\;
\pa_{ B } Y\right\rangle_{J}
 =  0   
\eeq
due to Eq.\ (\ref{I15}) for $\mu  =  {\rm const}$.
It follows then from Eqs.\ (\ref{W4}) and (\ref{W6}) that
\beq
\label{W7}
\left\langle J_{ A }\; \omega^{ A B }\; \pa_{ B } W\right\rangle_{J} =
- \frac{1}{2} J_{ A }\; \omega^{ A B } \; J_{ B }
 (-1)^{ \varepsilon_{ B } } .   
\eeq
Thus we have eliminated the average (\ref{I33}) of the gauge-fixing
master action $X$ from the new Ward identity (\ref{W7}). The price is that
we have got the non-homogeneity quadratic in the external sources $J$ in the
right-hand side in Eq.\ (\ref{W7}).

Finally, at the level of finite BRST-BV transformations, the
relation (\ref{I22}) yields
\beq
\label{W8}
\left\langle\exp\left\{\frac{i}{\hbar} J_{ A }\;\!( \overline{ z}^{ A
}-z^{ A })+ A \right\}\right\rangle_{J} = 1. 
\eeq
However, it is impossible to eliminate the average (\ref{I33})
of the gauge-fixing master action $X$ from (\ref{W8}).

\vspace{5mm}
\section{$W$-$X$ formulation to the $Sp(2)$-symmetric field-antifield formalism}

Let $z ^{ A }$ be the complete set of the variables necessary to the
$W$-$X$ formulation of the $Sp(2)$-symmetric field-antifield formalism
\cite{BLTl1,BLTl2,BLTl3}
\beq
\label{II1}
z^{ A }  =  \{ \Phi^{ \alpha }, \pi^{ \alpha a };
\Phi^{*}_{ \alpha a }, \Phi^{**}_{ \alpha } \} 
\eeq
whose Grassmann parities are
\beq
\label{II2}
\varepsilon( z^{ A } )  =  \{ \varepsilon_{ \alpha }, \varepsilon_{ \alpha } +1;
\varepsilon_{ \alpha } + 1, \varepsilon_{ \alpha } \}. 
\eeq
We denote the respective $z^{ A }$ derivatives as
\beq
\label{II3}
\pa_{ A }  =  \{ \pa_{ \alpha }, \pa_{ \alpha a };
\pa^{ \alpha a }_{*}, \pa^{ \alpha }_{**} \}. 
\eeq
Let $Z$ be the partition function:
\beq
\label{II4}
Z  = \int \! Dz D\lambda  \exp\left\{ \frac{i}{\hbar} [ W + X ] \right\} , 
\eeq
where $\lambda^{ \alpha }$ are Lagrange multipliers for gauge-fixing with
Grassmann parities
\beq
\label{II5}
\varepsilon( \lambda^{ \alpha } )  =  \varepsilon_{ \alpha }. 
\eeq
In the partition function (\ref{II4}),
the dynamical gauge-generating master action $W$ and the
gauge-fixing master action $X$ is defined to satisfy the respective quantum
master equation
\beq
\label{II6}
\left(\Delta^{ a }_{+}\exp\left\{ \frac{i}{\hbar} W \right\} \right)  = 0
\quad\Leftrightarrow\quad
\frac{1}{2} ( W, W )^{ a } + (V^{ a } W)  =  i \hbar (\Delta^{ a } W), 
\eeq
\beq
\label{II7}
\left(\Delta^{ a }_{-}\exp\left\{ \frac{i}{\hbar} X \right\} \right) = 0
\quad\Leftrightarrow\quad
\frac{1}{2}( X, X )^{ a } - (V^{ a } X)  =  i \hbar (\Delta^{ a } X). 
\eeq
In the above quantum master equations (\ref{II6}) and (\ref{II7}),
the $\Delta^{ a }, (\; ,\; )^{ a }$, $V^{ a }$ and $\Delta^{ a }_{\pm}$ are
the $Sp(2)$-vector-valued odd Laplacian
\beq
\label{II8}
\Delta^{ a } = \pa_{\alpha} \pa^{ \alpha a }_{*} (-1)^{ \varepsilon_{ \alpha } } +
\varepsilon^{ ab } \pa_{ \alpha b } \pa^{ \alpha }_{**} (-1)^{ \varepsilon_{ \alpha } +
1 }, 
\eeq
antibracket
\beq
\label{II9}
( f , g )^{ a } =(-1)^{\varepsilon_{ f } } [  [ \Delta^{ a } , f ], g ] 1
= f  \left[ \overleftarrow{\pa} _{ \alpha }~\overrightarrow{\pa}^{ \alpha a }_{*}
+\varepsilon^{ ab } \overleftarrow{\pa }_{ \alpha b }
~\overrightarrow{\pa }^{ \alpha }_{**}
\right]  g
- ( f \leftrightarrow g ) (-1)^{ ( \varepsilon_{ f } + 1 ) ( \varepsilon_{ g
} + 1 ) },                         
\eeq
special vector field
\beq
\label{II10}
V^{ a }  =  V^{A a }\partial_{A}
=  \varepsilon^{ ab }\;\! \Phi^{*}_{ \alpha b } \;\!\pa^{ \alpha }_{**}, \qquad
  ( V^{ a } z^{ A } ) = V^{A a } ,
\eeq
and
\beq
\label{II10b}
\Delta^{ a }_{\pm}:=\Delta^{ a } \pm \frac{i}{\hbar} V^{ a },
\eeq
respectively.
For the $W$-action, one should require that $W$ is independent of
$\pi^{ \alpha a }$,
\beq
\label{II11}
( \Phi^{**}_{ \alpha }, W ) = 0. 
\eeq
As to the $X$-action, it can be chosen in the form related to the gauge-fixing
Boson $F(\Phi)$,
\beq
\label{II12}
\nonumber
X  &=&  \Phi^{*}_{ \alpha a } \pi^{ \alpha a }
+ ( \Phi^{**}_{ \alpha } - F \overleftarrow{\pa}_{\alpha } ) \lambda^{ \alpha }  +
\frac{1}{2}F \overleftarrow{\pa }_{\alpha}~
\pi^{ \alpha a }\overleftarrow{\pa }_{\beta}
~\pi^{\beta b} \varepsilon_{ba}   \\
&=&
\Phi^{*}_{ \alpha a } \pi^{ \alpha a } + \Phi^{**}_{ \alpha } \lambda^{ \alpha }  +
\frac{1}{2} F \overleftarrow{d }^{ a } \overleftarrow{d }^{b} \varepsilon_{
ba }, 
\eeq
where
\beq
\label{II13}
\overleftarrow{d }^{ a }  =  \overleftarrow{\pa}_{ \alpha } \pi^{ \alpha a } -
\overleftarrow{\pa }_{ \alpha b } \lambda^{\alpha} \varepsilon^{ ba } 
\eeq
is a $Sp(2)$-vector-valued Fermionic differential that acts from the right.

In the integrand of the path integral (\ref{II4}),
consider now the following infinitesimal BRST transformation
\beq
\label{II14}
\delta z^{ A }
=  -\mu_{ a } (Y, z^{ A })^{ a }
- \frac{\hbar}{i}  ( \mu_{ a } ,z^{ A }  )^{ a }
+ 2 \mu_{ a }  V^{A a }
= -\frac{\hbar}{i} y^{-1}(y\mu_{ a }, z^{ A })^{ a }
+ 2 \mu_{ a } V^{A a },  
\eeq
where we have defined for later convenience
\beq
\label{II14b}
 Y~:=~X-W,\qquad
y~:=~\exp\left\{ \frac{i}{\hbar}Y \right\},
\eeq
and where $\mu_a=\mu_{ a }( z )$ is an infinitesimal $Sp(2)$ co-vector
valued Fermionic function. Its Jacobian has the form
\beq
\label{II15}
\ln J   = (-1)^{\varepsilon_{A}} (\partial_{A} \delta z^{ A })
=   ( Y , \mu_{ a } )^{ a } + 2 ( \Delta^{ a } Y)  \mu_{ a }  +
2 \frac{\hbar}{i} (\Delta^{ a }_{-} \mu_{ a }) .  
\eeq
The complete action in the partition function (\ref{II4}) transforms as
\beq
\label{II16}
\nonumber
\delta [ W + X ]  &=&  [ - ( W, W )^{ a } + ( X, X )^{ a } ] \mu_{ a }
+ \frac{\hbar}{i} ( W + X,  \mu_{ a } )^{ a } - 2[V^{ a } ( W + X)] \mu_{ a } \\
&=& 2 i  \hbar ( \Delta^{ a } Y ) \mu_{ a }
+ \frac{\hbar}{i} ( W + X,  \mu_{a} )^{ a }. 
\eeq
It follows from Eqs.\ (\ref{II15}) and (\ref{II16}) that
\beq
\label{II17}
\frac{i}{\hbar}\delta [ W + X ] + \ln J
=  2 (\sigma_{-}^{ a }( X ) \mu_{ a }), 
\eeq
where
\beq
\label{II18}
(\sigma_{-}^{ a } ( X ) f)
= ( X, f )^{ a } + \frac{\hbar}{i}( \Delta^{ a }_{-} f) 
\eeq
is the $Sp(2)$ vector-valued quantum BRST generator.

The Eq.\ (\ref{II17}) tells us that the BRST transformation (\ref{II14})
induces the following variation
\beq
\label{II19}
\delta X  =  2 \frac{\hbar}{i} (\sigma_{-}^{ a } ( X )  \mu_{ a }). 
\eeq
to the $X$-action in the integrand of the path integral (\ref{II4}).
We conclude that the partition function (\ref{II4}) and the quantum master
equation (\ref{II7}) for $X$ are both stable under the infinitesimal variation
(\ref{II19}).

Next let $t$ be a Bosonic parameter. It is natural to define a one-parameter
subgroup $t\mapsto \overline{z}^{ A }(t)$ of finite BRST transformations
by the Lie equation
\beq
\label{II20}
\frac{d\overline{z}^{A}}{dt} = \overline{\mathcal{H}}^{A},
\qquad \overline{z}^{A} |_{  t=0 }   =  z^{ A }, 
\eeq
where
\beq
\label{II32}
\mathcal{ H }  =  \mathcal{ H }^{A} \partial_{A}
=  -  \mu_{ a }  {\rm ad}^{ a }(Y )
-\frac{ \hbar }{ i } {\rm ad}^{ a }( \mu_{ a } )
+  2 \mu_{ a }V^{ a }
=  -\frac{ \hbar }{ i }y^{-1} {\rm ad}^{ a }(y \mu_{ a } )
+  2 \mu_{ a }V^{ a } , 
\eeq
is the corresponding vector field with components
\beq
\label{II30b}
\mathcal{ H }^{A} := -\mu_{ a } (Y, z^{ A } )^{ a }
- \frac{\hbar}{i}(\mu_{ a }, z^{ A } )^{ a } +  2 \mu_{ a }V^{A a }
= -\frac{\hbar}{i} y^{-1}(y\mu_{ a }, z^{ A })^{ a } + 2 \mu_{ a }V^{A a } .
\eeq
This equation implies the $Sp(2)$-vector-valued counterpart to
the equation (\ref{I18a}),
\beq
\nonumber
\label{II20a}
 \left(  \frac{ d }{ dt } - 2 \overline{\mu}_{ a }\overline{V }^{ a } \right)
( \overline{y}\overline{\mu }_{ b } )
&=&\left( \frac{d\overline{z}^{A}}{dt}
- 2  \overline{\mu}_{ a }\overline{V }^{Aa }  \right)
\underline{\partial}_{A}(\overline{y} \overline{\mu}_{ b })  \\
= -\frac{\hbar}{i} \overline{y}^{-1}\overline{(y\mu_{a},z^{A})^{a}}~
\underline{\partial}_{A}(\overline{y} \overline{\mu}_{ b })
&=&  -\frac{\hbar}{i} \overline{y}^{-1}\overline{(y\mu_{ a },y\mu_{ b })^{ a }},
\eeq
which cannot be completely integrated explicitly to yield a
counterpart to the conservation law (\ref{I18b}).

The Jacobian of the transformation (\ref{II20}) satisfies the equation
\beq
\label{II21}
\frac{d\ln J}{dt} = \overline{\mathrm{div}\mathcal{H}},
\qquad \mathrm{div}  \mathcal{H}
:=  (-1)^{\varepsilon_{A}} \partial_{A} \mathcal{H}^{A}
=  ( Y, \mu_{ a } )^{ a }  + 2  (\Delta Y^{ a } ) \mu_{ a }
+ 2 \frac{\hbar}{i} (\Delta^{ a }_{-} \mu_{ a } ) .   
\eeq
The complete action in Eq.\ (\ref{II4}) satisfies the equation
\beq
\nonumber
\label{II22}
\frac{d}{dt}[\overline{W } + \overline{X} ]
&=& \frac{d\overline{z}^{A}}{dt}~ \underline{\partial}_{A}
\left[\overline{W } + \overline{X} \right]
= \left[-\overline{\mu}_{a} \overline{(Y, z^{ A } )^{a}}
 - \frac{\hbar}{i}\overline{(\mu_{a}, z^{ A } )^{a}}
+2\overline{\mu}_{a} \overline{V}^{Aa} \right]
\underline{\partial}_{A} \left[\overline{W } + \overline{X} \right]  \\
&=& \left[ - \overline{( W , W )^{a}}
+  \overline{( X , X )^{a}}   \right] \overline{\mu }_{a}   +
\frac{\hbar}{i}  \overline{(W +X , \mu _{a} )^{a}}
- 2 [\overline{V}^{ a } ( \overline{W } + \overline{X } )] \overline{\mu}_{a} .
\eeq
It follows from Eqs.\ (\ref{II21}) and (\ref{II22}) that
\beq
\label{II23}
\frac{d}{dt}\left[\frac{i}{\hbar}(\overline{W } + \overline{X})
+  \ln J \; \right]  = \overline{a},
\eeq
where we have defined for later convenience
\beq
\label{IIa}
a:= 2 (\sigma_{-}^{ a } (X )\mu_{ a }).   
\eeq
By integrating within $0\leq t\leq 1$, we get from Eq.\ (\ref{II23})
\beq
\label{II24}
\overline{W } + \overline{X }   +  \frac{\hbar}{i}  \ln J
 = W + X  +\frac{\hbar}{i} A,   
\eeq
where we have defined the average
\beq
\label{II31}
\qquad A:= \int_{0}^{1} \! dt ~ \overline{ a }
=  \int_{0}^{1} \! dt~\left(e^{ t \mathcal{ H } }a\right)
= E( \mathcal{ H } )a .
\eeq
The Eq.\ (\ref{II24}) shows the finite effective change in $X$ induced by
the finite transformation $z^{ A } \rightarrow \overline{ z}^{ A }$ in
Eq.\ (\ref{II4}).

Now consider the left-hand side $\overline{\mathcal{Y}}^{ a } $ of the
transformed quantum master equation (\ref{II7}), where
\beq
\label{II25}
\mathcal{Y}^{ a }
:= \frac{1}{2} (X,X )^{ a } +\frac{\hbar}{i}(\Delta^{ a }_{-} X).   
\eeq
We have the following Cauchy initial value problem
\beq
\label{II26}
\frac{d\overline{\mathcal{Y}}^{ a }}{dt}
= (\overline{\mathcal{H}}~\overline{\mathcal{Y}}^{ a })
\quad\wedge\quad
\overline{\mathcal{Y} }^{ a }|_{ t = 0 }=0
\quad\Rightarrow\quad\overline{\mathcal{Y} }^{ a } \equiv 0
\eeq
for arbitrary $t$.

Thereby, we have confirmed that the quantum master equation (\ref{II7}) is
stable under the finite BRST-BV transformation generated by Eq.\
(\ref{II20}). Of course, the general expression (\ref{II4}) itself is stable
under the same transformations, as well.

The $Sp(2)$-extended quantum master equation
\beq
\label{II27}
\left( \Delta^{ a }_{-} \exp\left\{\frac{i}{\hbar} X' \right\}\right)  =  0
\eeq
for the new gauge-fixing master action,
\beq
\label{II28}
X'  =  X  + \frac{\hbar}{i}A,   
\eeq
must be interpreted similarly to what we have explained in Section 2 with
respect to the Eq.\ (\ref{I25}).
For instance, the $Sp(2)$ vector valued counterpart to the Eq.\
(\ref{I34}) reads
\beq
\label{II30}
(\sigma_{-}^{ a }( X ) \exp\{ A \})  = 0  \quad\Leftrightarrow\quad
\frac{ \hbar }{ 2 i } ( A, A )^{a} + (\sigma_{-}^{ a }( X ) A)   =  0. 
\eeq
Finally, the respective $Sp(2)$ symmetric counterpart to the Eqs.\
(\ref{I49})-(\ref{I50}) reads
\beq
\label{II33}
\exp\left\{ \frac{ i }{ \hbar }  X' \right\}
&=&\left(\exp\left\{\frac{ 1 }{ 2 }  \varepsilon_{ab}
\left[ \Delta^{ b }_{-} , \left[  \Delta^{ a }_{-} , \nu \right] \right]  \right\}
\exp\left\{ \frac{ i }{ \hbar }  X \right\}\right) \\
&=&\exp\left\{ \frac{ i }{ \hbar }  X \right\}  +
\left(\frac{ 1 }{2} \varepsilon_{ab}  \Delta^{ b}_{-} \Delta^{ a }_{-} \nu
E\left(  \frac{ 1 }{ 2 }  \varepsilon_{cd}
\left[   \Delta^{ d }_{-} , \left[  \Delta^{ c }_{-} , \nu \right] \right] \right)
\exp\left\{ \frac{ i }{ \hbar }  X \right\}\right), \quad 
\label{II34}
\eeq
with $\nu$ being any finite Bosonic operator.

\vspace{5mm}
\section{Ward identities in the $Sp(2)$-extended $W$-$X$ formulation}

Let $J_{ A }$ be external sources to the variables $z^{ A }$; then the
integral (\ref{II4}) generalizes to the generating functional
\beq
\label{WS1}
Z [ J ] = \int  Dz D\lambda
\exp\left\{ \frac{i}{\hbar}[\; W  +  X +  J_{ A }  z^{ A} \;]\right \}.
\eeq
Arbitrary variation $\delta z^{ A }$ yields the equations of motion,
\beq
\label{WS2}
\left\langle \pa_{ B } ( W + X )\right\rangle_{J}
+  J_{ B } (-1)^{ \varepsilon_{ B } }
  =  0, 
\eeq
where $\langle \ldots \rangle_{J}$ is the source-dependent mean-value
\beq
\label{WS3}
\langle \ldots \rangle_{J}   = \frac{1}{Z[J]}
\int  Dz  D\lambda  ( \ldots )  \exp\left\{ \frac{i}{\hbar} [\;  W  +  X
+  J_{ A } z^{ A }\; ] \right\}.        
\eeq
It follows from Eq.\ (\ref{WS2}) that
\beq
\label{WS4}
\left\langle J_{ A } \; \omega^{ A B a } \;
\pa_{ B } ( W + X )\right \rangle_{J}
+ J_{ A }\; \omega^{ A B a }\; J_{ B } (-1)^{ \varepsilon_{ B } }   = 0,
\eeq
where
\beq
\label{WS5}
\omega^{ A B a }  =  ( z^{ A },  z^{ B } )^{a}  = {\rm const}     
\eeq
is the fundamental $Sp(2)$ antibracket. In Eq.\ (\ref{WS1}), the BRST-BV
variation (\ref{II14}) yields
\beq
\label{WS6}
\left\langle  J_{ A }  [ \;\omega^{ A B a }\; \pa_{ B }Y
- 2  V^{A a }  (-1)^{\varepsilon_{ A } } ] \right\rangle_{J}   =   0,   
\eeq
due to Eq.\ (\ref{II17})  for $\mu_{ a }  =  {\rm const}$.
It follows then from Eqs.\ (\ref{WS4}) and (\ref{WS6}) that
\beq
\label{WS7}
\left\langle  J_{ A }  [\; \omega^{ A B a } \;\pa_{ B } W
+  V^{ Aa } (-1)^{\varepsilon_{ A } } ]\right \rangle_{J}   =
- \frac{1}{2} J_{ A }\; \omega^{ A B a }\; J_{ B }
(-1)^{ \varepsilon_{ B } } .   
\eeq
Thus we have eliminated the average (\ref{II31}) of the gauge-fixing
master action $X$ from the new Ward identity (\ref{WS7}).
The price is that we have got the non-homogeneity quadratic
in the external sources $J$ in the right-hand side in Eq.\ (\ref{WS7}).

Finally, at the level of finite BRST-BV transformations, the
relation (\ref{II24}) yields \beq \label{WS8} \left\langle
\exp\left\{\frac{i}{\hbar} J_{ A }(\overline{z}^{ A } - z^{ A } ) +
A \right\} \right\rangle_{J}  = 1.  
\eeq
However, it is impossible to eliminate the average (\ref{II31}) of the
gauge-fixing master action $X$ from Eq.\ (\ref{WS8}).

\vspace{5mm}
\section{Conclusions}

Notice that, on one hand (and in contrast to the original $Sp(2)$-formulation 
\cite{BLTl1,BLTl2,BLTl3}), in the $Sp(2)$-symmetric $W$-$X$ formulation, 
the anti-canonical dynamical activity of the variables
$\{\pi^{ \alpha a }, \Phi^{**}_{ \alpha }\}$ \cite{BM}, as represented by the
second term in Eq.\ (\ref{II8}) and in the square bracket of Eq.\ (\ref{II9}), 
is of crucial importance to satisfy the  quantum master equation (\ref{II7})
with the anzatz (\ref{II12}) for $X$. 
On the other hand, $\pi^{ \alpha a}$ and $\Phi^{**}_{\alpha}$ are kept as 
dynamically passive (antibracket-commuting) variables in the $W$-action.
Thus, one may realize what is the price of coexistence between 
the $Sp(2)$-symmetry and the complementary $W$-$X$ duality of the 
quantum master equations (\ref{II6}) and (\ref{II7}).

\vspace{5mm}
\section*{Acknowledgments}
\noindent
The work of I.A.B.\ is supported in part by the RFBR grants 14-01-00489 and
14-02-01171. The work of K.B.\ is supported by the Grant Agency of the Czech
Republic (GACR) under the grant P201/12/G028. The work of P.M.L.\ is partially
supported by the Presidential grant 88.2014.2 for LRSS and by the RFBR grant
15-02-03594.

\appendix
\vspace{5mm}
\section{Algebra of the $\sigma$-operators}

In this Appendix we present the general formal algebra of the
$\sigma$-operators, both in the standard and the $Sp(2)$ case.

In the standard case we introduce the $\sigma$-operator 
\beq
\label{A1}
\sigma( F )  :=   \frac{ \hbar }{ i }  \exp\left\{
- \frac{ i }{ \hbar } F \right\} \Delta   
\exp\left\{ \frac{ i }{ \hbar } F \right\}
=  \frac{ \hbar }{ i }\Delta + {\rm ad}(F) 
+\left(\Delta F + \frac{i}{2\hbar}(F,F)\right)  
\eeq
for any Bosonic functional $F$. It inherits the nilpotency
\beq 
\label{A1nilp}
 \Delta^{2}=0\qquad\Rightarrow \qquad  (\sigma( F ) )^{ 2 }  = 0.
\eeq
Then straightforward calculation gives the following results for
the commutator of $\sigma(W)$ and $\sigma(X)$
\beq
\label{A2}
[  \sigma( W ),  \sigma( X ) ] = {\rm ad} (C),
\eeq
where
\beq
\label{A2C}
C := \frac{\hbar }{ i } (\Delta(W+X)) +(W,X)
=  - \frac{ 1 }{ 2 } ( Y, Y )
=  \frac{ 4 \hbar }{ i } ( \sigma( \frac{ W+X }{ 2 }  ) 1 ) , 
\eeq
and where the quantum master equations for $W$ and $X$ are used.

In the $Sp(2)$ case the set of operators $\sigma^{a}(F)$, $\sigma^{a}_{\pm}(F)$ 
for any Bosonic functional $F$ is introduced
\beq
\label{A5}
\sigma^{ a } ( F )  :=  \frac{ \hbar }{ i }  \exp\left\{ - \frac{ i }{
\hbar } F \right\}  \Delta^{ a }  \exp\left\{ \frac{ i }{ \hbar } F \right\}
=  \frac{ \hbar }{ i }\Delta^{ a } + {\rm ad}^{ a }(F) 
+\left(\Delta^{ a } F + \frac{i}{2\hbar}(F,F)^{ a }\right) , 
\eeq

\beq
\label{A3}
\sigma^{ a }_{\pm} ( F )  :=  \frac{ \hbar }{ i }  \exp\left\{ - \frac{ i
}{ \hbar } F \right\} \Delta^{ a }_{\pm}  
\exp\left\{ \frac{ i }{ \hbar } F \right\}
=  \frac{ \hbar }{ i }\Delta_{\pm}^{ a } + {\rm ad}^{ a }(F) 
+\left(\Delta_{\pm}^{ a } F + \frac{i}{2\hbar}(F,F)^{ a }\right) .
\eeq
The $Sp(2)$ nilpotency reads
\beq 
\nonumber
\label{A3nilp}
&& [\Delta^{a},\Delta^{b}]=0,\quad  [\Delta^{\{a},V^{b\}}]=0,
\quad [V^{a},V^{b}]=0,\\
\quad&\Rightarrow& \quad  
[\sigma^{a}( F ),\sigma^{b} (F)]  = 0, \quad 
[\sigma_{\pm}^{a}( F ),\sigma_{\pm}^{b} (F)]  = 0.
\eeq
Taking into account the quantum master equations for $W$ and $X$
from Eqs.\ (\ref{A5})-(\ref{A3}) it follows that
\beq
\label{A8}
[ \sigma^{ \{ a }_{+}( W ),  \sigma^{ b \} }_{-}( X ) ] 
=  {\rm ad}^{ \{ a } (C^{ b \} }  ),
\eeq
where
\beq
\label{A8C}
C^{b} := \frac{\hbar }{ i } (\Delta_{-}^{b} W) 
+\frac{\hbar }{ i } (\Delta_{+}^{b} X) + (W,X)^{b}
= - \frac{ 1 }{ 2 } ( Y, Y )^{ b  }  +  2 ( V^{ b  } Y ) 
=  \frac{ 4 \hbar }{ i }  ( \sigma^{ b } ( \frac{ W+X }{ 2 } ) 1 ) .
\eeq

\end{document}